\documentclass{article}
\usepackage{amsmath}
\usepackage{graphicx}
\usepackage{eurosym}
\usepackage{soul}
\usepackage{color}
\setulcolor{red}

\begin{document}

\title{Modelling the Earth's magnetic field}
\author{}
\author{Nuno Barros e S\'a$^1$, Louren\c co Faria$^2$,\\ Bernardo Alves$^2$, Miguel Cymbron$^2$}

\date{$^1$ Institute of Marine Sciences - Okeanos, University of the Azores, Rua Prof. Doutor Frederico Machado 4, 9901-862 Horta, Portugal\\
$^2$ Escola Secund\'aria Domingos Rebelo, Av. Antero de Quental, 9504-501 Ponta Delgada, Portugal}
\maketitle

\begin{abstract}
The Earth's magnetic field can be decomposed into spherical harmonics, and the exact coefficients of the decomposition can be determined through a few measurements of its value at different locations. Using measurements from a magnetometer on board the International Space Station, we computed the multipolar moments of an expansion in spherical harmonics to order 3, calculated the location of the magnetic dip poles, and produced an intensity map of the magnetic field across the globe. The accuracy of the results was evaluated by comparison with data from the International Geomagnetic Reference Field.
\end{abstract}

\section{Introduction}

Laplace equation
\begin{equation}
\nabla ^{2}V=0
\end{equation}
shows up early in an Electromagnetism course, in the context of Electrostatics (\cite{jac}, Ch.1). Solving the Laplace equation in spherical coordinates leads to the expansion of
the electrostatic potential $V$ in spherical harmonics \cite{jac}:%
\begin{equation}
V\left( r,\theta ,\phi \right) =\sum_{l=1}^{\infty }\sum_{m=0}^{l}\left[
A_{lm}r^{l}+\frac{B_{lm}}{r^{l+1}}\right] \sqrt{\frac{\left( 2l+1\right)
\left( l-m\right) !}{4\pi \left( l+m\right) !}}P_{lm}\left( \cos \theta
\right) e^{im\phi }\ .  \label{eu3}
\end{equation}%
Here $r$, $\theta$ and $\phi$ are respectively the distance to the origin, the polar angle and the azimuthal angle. $A_{lm}$ and $B_{lm}$ are the expansion coefficients, and $P_{lm}\left( x\right) $ are the associated Le\-gendre polynomials; $l$ and $m$ are called the degree and the order of the spherical harmonics respectively.

In this article, we propose a real-life application of the multipolar expansion of the solution to Laplace equation to the case of the Earth's magnetic field. From the comparison of a limited number of measurements to this series expansion, one can reconstruct the Earth's magnetic field above its surface and determine the location of the magnetic dip poles. This project relates the theory with real data, and allows easy comparison with tabulated data, making it a nice tool in the teaching of Electromagnetism.

In the work carried out by our students (presented in Section \ref{s5}), the measurements of the Earth's magnetic field were performed on board the International Space Station (ISS) at points along its orbit. However, this project can be applied to other situations, such as measurements collected from different locations on the surface of the Earth (Section \ref{s4}). In addition to giving students the thrill of working with data collected on the ISS, using these measurements challenged students to understand not only electromagnetism but also reference frame transformations. Moreover, the comparison between theory and experiment required the construction of algorithms and some amount of programming. Using knowledge acquired in other disciplines raised students' awareness that, to tackle practical problems, they need to interconnect different disciplines of their curriculum.

\section{Multipolar expansion of the magnetic field on the surface of the
Earth}\label{sss}

The Earth's magnetic field is complex and evolves in time.  Deriving tractable models to describe it as accurately as possible is of practical importance for navigation \cite{bar}, defense, and land surveying, and also of scientific importance for study of topics ranging from properties of the Earth's core \cite{sp} to animal migration (see \cite{md} for a review).

The intensity of the Earth's magnetic field along its surface varies between 20 - 65 $\mu$T. This field can be described, to a first approximation, by a magnetic dipole placed at the center of the Earth and tilted by $11^{\circ}$ with respect to Earth's rotation axis. The geomagnetic poles are the (antipodal) points of intersection of the dipole axis with the surface of the Earth.

The real magnetic field of the Earth is, however, not perfectly dipolar. For example, the intensity of the magnetic field is abnormally low in the South Atlantic (the so-called South Atlantic anomaly), and the dip poles (points on the surface of the Earth where the magnetic field is vertical) do not coincide with the geomagnetic poles, neither are they antipodal (on opposite points on the Earth's surface). Improved descriptions of the Earth's magnetic field can be obtained by going to higher orders of a multipolar expansion, as we shall see below.

We start from Amp\`ere-Maxwell's law
\begin{equation}
\vec{\nabla}\times \vec{H} =\vec{j}+\frac{\partial\vec{D}}{\partial t}\ ,\label{noo}
\end{equation}%
where $\vec{H}$ is the magnetic field strength and $\vec{j}$ is the free current density. One can neglect the time-variation of the displacement field $\vec{D}$: $\partial\vec{D}/\partial t=0$, since the Earth's magnetic field varies slowly in time. At the surface of the Earth and in the atmosphere, one can further set $\vec{j}=0$ because the Earth's magnetic field is essentially produced by currents inside its core \cite{mer}. Moreover, in the atmosphere, $\vec{H}=\vec{B}/\mu_0$, where $\vec{B}$ is the magnetic field and $\mu _{0}$ is the magnetic permeability of the vacuum. Equation (\ref{noo}) then simplifies to
\begin{equation}
\vec{\nabla}\times \vec{B} =0\quad \Rightarrow \quad \vec{B} =-\vec{\nabla}\psi\ , \label{eu5}
\end{equation}
where $\psi$ is called the scalar magnetic potential. Plugging Eq. (\ref{eu5}) in Gauss's law for the magnetic field, one gets
\begin{equation}
\vec{\nabla}\cdot \vec{B} =0\quad \Rightarrow \quad \nabla ^{2}\psi =0\ ,
\end{equation}
that is, the scalar potential $\psi$ obeys Laplace's equation.

The scalar magnetic potential can therefore be expanded into spherical coordinates, in the same way as in Eq. (\ref{eu3}). It is natural to choose the center of the Earth as the origin of the coordinates and the axis of rotation of the Earth as the polar axis. The terms with coefficients $A_{lm}$ grow with $r$, and hence cannot represent the magnetic field created by currents circulating inside the Earth. We shall therefore set them to zero, as, on the surface of the Earth and in its atmosphere, the influence of external sources is not generally significant. Moreover, since there are no magnetic charges, the monopolar term is $B_{00}=0$. Finally, the terms with coefficients $B_{lm}$ can be rearranged to give:
\begin{equation}
\psi =a\sum_{l=1}^{N}\left( \frac{a}{r}\right) ^{l+1}\sum_{m=0}^{l}\left[
g_{l}^{m}\cos \left( m\phi \right) +h_{l}^{m}\sin \left( m\phi \right) %
\right] P_{l}^{m}\left( \sin \theta \right)\ ,\label{bleu}
\end{equation}%
where:
\begin{itemize}
\item[--] The angle $\theta $ is measured from the equator and is positive in the northern hemisphere. It therefore represents (geocentric) latitude and not colatitude.
\item[--] Distances are normalized to the mean Earth's radius $a=6.371\,2\times 10^{6}$ m.
\item[--] The model contains $N\left( N+2\right) $ parameters: the coefficients $g_{l}^{m}$ and $h_{l}^{m}$ (Note that $h_l^0=0$ for all $l$). Since, for practical calculations, the expansion cannot contain infinite terms, we truncated the expansion to order $N=3$ in $l$.
\item[--] We chose the Schmidt quasi-normalization for the associated Legendre polynomials, $P_{l}^{m}$, which is normally used in Geomagnetism \cite{mer,al}, that is,
    \begin{equation}
    P_{l}^{m}=\sqrt{\frac{2\left( l-m\right) !}{\left( 1+\delta _{0}^{m}\right)\left( l+m\right) !}}P_{lm}\ ,
    \end{equation}
    where $\delta$ is the Kronecker delta and
    \begin{equation}
    P_{lm}\left( x\right) =\frac{1}{2^{l}l!}\left( 1-x^{2}\right) ^{m/2}\frac{d^{l+m}}{d^{l+m}x}\left( x^{2}-1\right) ^{l}\ .
    \end{equation}
\end{itemize}

Concerning this last point, we should point out that conventions regarding the normalization and sign (the Condon-Shortley phase) of the associated Legendre polynomials vary in different areas of physics. The values for coefficients $g_{l}^{m}$ and $h_{l}^{m}$ then depend on the preferred convention. We chose to express the multipolar expansion to order $N=3$ obtained with this normalization as:
\begin{eqnarray}
\psi &=&\frac{a^{3}}{r^{2}}\left\{ g_{1}^{0}\sin \theta +\left[
g_{1}^{1}\cos \phi +h_{1}^{1}\sin \phi \right] \cos \theta \right\} +  \notag
\\
&&+\frac{a^{4}}{r^{3}}\left\{ g_{2}^{0}\left( 1-\frac{3}{2}\cos ^{2}\theta
\right) +\sqrt{3}\left[ g_{2}^{1}\cos \phi +h_{2}^{1}\sin \phi \right] \sin
\theta \cos \theta +\right.  \notag \\
&&\left. +\frac{\sqrt{3}}{2}\left[ g_{2}^{2}\cos \left( 2\phi \right)
+h_{2}^{2}\sin \left( 2\phi \right) \right] \cos ^{2}\theta \right\} +
\notag \\
&&+\frac{a^{5}}{r^{4}}\left\{ g_{3}^{0}\sin \theta \left( 1-\frac{5}{2}\cos
^{2}\theta \right) +\sqrt{\frac{3}{2}}\left[ g_{3}^{1}\cos \phi
+h_{3}^{1}\sin \phi \right] \times \right.  \notag \\
&&\times \cos \theta \left( 2-\frac{5}{2}\cos ^{2}\theta \right) +\sqrt{%
\frac{15}{2}}\left[ g_{3}^{2}\cos \left( 2\phi \right) +h_{3}^{2}\sin \left(
2\phi \right) \right] \sin \theta \cos ^{2}\theta +  \notag \\
&&\left. +\sqrt{\frac{15}{8}}\left[ g_{3}^{3}\cos \left( 3\phi \right)
+h_{3}^{3}\sin \left( 3\phi \right) \right] \cos ^{3}\theta \right\}\ .
\label{eu6}
\end{eqnarray}

With this in mind, the magnetic field can be calculated from Eqs. (\ref{eu5}) and (\ref{eu6}):%
\begin{equation}
\vec{B}=-\vec{\nabla}\psi =-\left( \frac{\partial \psi }{\partial r}\hat{e}%
_{r}+\frac{1}{r}\frac{\partial \psi }{\partial \theta }\hat{e}_{\theta }+%
\frac{1}{r\cos \theta }\frac{\partial \psi }{\partial \phi }\hat{e}_{\phi
}\right)\ . \label{eu7}
\end{equation}

Let us note that the multipolar expansion in lowest order, $N=1$, leads to the familiar formula%
\begin{equation}
\vec{B}\left( \vec{r}\right) =\frac{\mu _{0}}{4\pi }\left[ \frac{3\left(
\vec{m}\cdot \vec{r}\right) \vec{r}}{r^{5}}-\frac{\vec{m}}{r^{3}}\right]\ ,
\end{equation}%
where the magnetic dipole moment is given by
\begin{equation}
\vec{m}=\frac{4\pi a^{3}}{\mu _{0}}\left(
g_{1}^{1},h_{1}^{1},g_{1}^{0}\right)\ .
\end{equation}%

Equations (\ref{bleu}) and (\ref{eu7}) are in fact those used in the International Geomagnetic Reference Field (IGRF), with $N=13$ for the latest (2020) model \cite{al}. The IGRF computes the values of the multipolar moments $g_{l}^{m}$ and $h_{l}^{m}$ from large sets of data obtained from observatories and satellites across the globe \cite{swarm}. It is updated every 5 years, and, for each release, contains both the multipolar moments and their expected (linear) time variation for the next 5-year period \cite{al}. In the following, we will compare our results with the much more precise results of the IGRF. We chose the IGRF as reference, as we are more familiar with it, the other major model of the Earth's magnetic field using a spherical harmonics expansion, with $N=12$, being the WMM (World Magnetic Model) \cite{wmm}.

A note should be made about the comparison of the $N=13$ IGRF moments to our $N=3$ results. If we had access to measurements covering the entire surface of the Earth, then a model with $N=13$ should produce the same moments up to order 3 as a model with $N=3$. But that is not the case with a finite number of measurements (for a clear discussion of this issue, see \cite{mer}, Chap. 2.3), and the less well-spaced the measurements, the bigger the difference is in the moments evaluated by the different order models. However, the IGRF model is based on a large and well-spaced set of measurements, and the absolute value of the moments decrease fast with increasing order. This makes the values of the lowest order moments in the IGRF model almost insensitive to further raising the order of the expansion, so that they can be considered as good estimates of the real values of the moments. The choice of the $N=3$ order for the truncation of the series in our model was made to make the problem interesting enough for the students, while still rendering it possible to solve using a personal computer in a reasonable amount of time.

\section{Calculating the multipolar moments from\break magnetometer readings}\label{s4}

We would now like to show how measurements of the magnetic field on different locations around the globe, either at its surface or in the atmosphere, enables the determination of the multipolar moments $g_{l}^{m}$ and $h_{l}^{m}$ via Eqs. (\ref{eu6}) and (\ref{eu7}). These equations are written in a spherical geocentric frame (where we may assume that the polar axis is the rotation axis of the Earth, and that the azimuthal angle measures longitude), while magnetometer readings typically come in the cartesian frame of the magnetometer, so that, in order to compare the two, one has to convert between the two frames at each location.

We begin by writing Eq. (\ref{eu7}) in the cartesian Earth-centered Earth-fixed frame (ECEF) $(\hat{e}_{x},\hat{e}_{y},\hat{e}_{z})$, which is such that
\begin{eqnarray}
\hat{e}_{r} &=&\cos \theta \cos \phi\, \hat{e}_{x}+\cos \theta \sin \phi\, \hat{e%
}_{y}+\sin \theta\, \hat{e}_{z} \label{ma1}\\
\hat{e}_{\theta } &=&-\sin \theta \cos \phi\, \hat{e}_{x}-\sin \theta \sin \phi\,
\hat{e}_{y}+\cos \theta\, \hat{e}_{z} \label{ma2}\\
\hat{e}_{\phi } &=&-\sin \phi\, \hat{e}_{x}+\cos \phi\, \hat{e}_{y}\ .\label{ma3}
\end{eqnarray}
Note that we used lowercase letters for the cartesian axes of the ECEF frame, in order to differentiate them from those of the magnetometer, for which we shall use capital letters.

\begin{figure}[tbp]
\centering
\includegraphics[width=.4\textwidth]{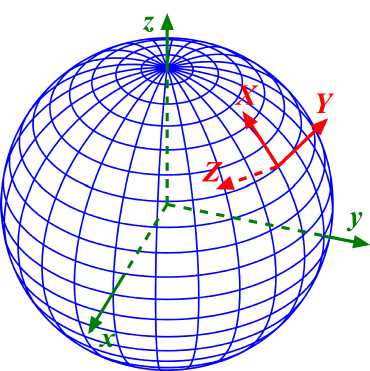}
\caption{The cartesian Earth-centered Earth-fixed frame (ECEF, lowercase letters) and the local frame of the magnetometer (uppercase letters).}
\label{glob}
\end{figure}

For the purpose of a student project, readings of the magnetic field across the globe are freely available from the global scientific magnetic observatory network, through the Intermagnet website \cite{inter}. Alternatively, a collaboration can be set up among schools using lower precision magnetometers, which are available at low cost, as was done in \cite{beg}.

For most magnetic measurements \textit{on the Earth's surface}, the magnetometer's frame is chosen such that the $X$ axis of the magnetometer points towards the north, its $Y$ axis points towards the east and its $Z$ axis points towards the center of the Earth (Fig. \ref{glob}). It is an instructive task for students to obtain the equations that relate the components of the field in the magnetometer frame, $B_{X}$, $B_{Y}$ and $B_{Z}$, to its components in the ECEF frame $B_{x}$, $B_{y}$ and $B_{z}$:
\begin{eqnarray}
B_{X} &=&-\cos \phi \sin \theta_{\rm d} B_{x}-\sin \phi \sin \theta_{\rm d} B_{y}+\cos
\theta_{\rm d} B_{z}\label{nu1} \\
B_{Y} &=&-\sin \phi B_{x}+\cos \phi B_{y} \\
B_{X} &=&-\cos \phi \cos \theta_{\rm d} B_{x}-\sin \phi \cos \theta_{\rm d} B_{y}-\sin
\theta_{\rm d} B_{z}\ ,\label{nu3}
\end{eqnarray}
where $\theta_{\rm d}$ stands for the geodetic latitude, which is the angle made by the normal to the surface of the Earth with the plane of the equator. Geocentric latitude $\theta$, on the other hand, is the angle made by the equator with a straight line connecting the location with the center of the Earth. The two quantities differ by a very small amount but they are not identical, since the Earth is not perfectly spherical. The World Geodetic System (WGS84) uses a reference ellipsoid for the surface of the Earth defined by the equation
\begin{equation}
\frac{x^{2}+y^2}{A^{2}}+\frac{z^2}{B^{2}}=1\label{re}
\end{equation}
where, by definition, $A=6\, 378\, 137$ m and $A/\left( A-B\right) =298.257\,223\,563$ exactly, from which $B\approx 6\, 356\, 752.314\, 245$ m follows. The relation between geocentric latitude $\theta$ and geodetic latitude $\theta_{\rm d}$ is given by \cite{hof}
\begin{equation}
A^{2}\tan\theta=B^{2}\tan\theta_{\rm d}\ .\label{reu}
\end{equation}
If the location of the magnetometers is known in terms of elevation $h$, longitude and geodetic latitude, as it normally is, one can convert it to the cartesian coordinates of the ECEF by \cite{hof}
\begin{eqnarray}
x &=&\left[ A^2/s\left(\theta_{\rm d}\right) + h \right] \cos \theta_{\rm d} \cos \phi\label{piu1} \\
y &=&\left[ A^2/s\left(\theta_{\rm d}\right) + h \right] \cos \theta_{\rm d} \sin \phi \\
z &=&\left[ B^2/s\left(\theta_{\rm d}\right) + h \right] \sin \theta_{\rm d}\ ,\label{piu2}
\end{eqnarray}
where
\begin{equation}
s\left(\theta_{\rm d}\right) =\sqrt{A^{2}\cos^2\theta_{\rm d}+B^{2}\sin^2\theta_{\rm d}}\ .
\end{equation}

Once the measurements have been performed, the experimental values can be compared to Eqs. (\ref{nu1})-(\ref{nu3}) to derive the value of the multipolar moments. This can be done using the method of least squares. Let $b_{X}^{i},b_{Y}^{i}$ and $b_{Z}^{i}$ be the three components of the magnetic field measured in the magnetometer's frame at location $i$, and $B_{X}^{i},B_{Y}^{i}$ and $B_{Z}^{i}$ the corresponding values predicted by Eqs. (\ref{eu6}) and (\ref{nu1})-(\ref{nu3}). The multipolar moments can be found by minimizing the function
\begin{equation}
S\left[ g_{l}^{m},h_{l}^{m}\right] =\sum_{i=1}^{P}\left[ \left( B_{X}^{i}%
\left[ g,h\right] -b_{X}^{i}\right) ^{2}+\left( B_{Y}^{i}\left[ g,h\right]
-b_{Y}^{i}\right) ^{2}+\left( B_{Z}^{i}\left[ g,h\right] -b_{Z}^{i}\right)
^{2}\right]\ , \label{blaba}
\end{equation}%
with respect to the $g_{l}^{m}$ and $h_{l}^{m}$. Here, $g$ denotes all $g_{l}^{m}$, and $h$ all $h_{l}^{m}$. $P$ is the number of data points.

After finding the best fit for $g_{l}^{m}$ and $h_{l}^{m}$, one can put the obtained values back in Eq. (\ref{eu6}) and compute the magnetic field in any other location $\vec B\left( r,\theta ,\phi\right)$. One can even find the location of the Earth's magnetic dip poles by minimizing, for the north and south poles respectively, the functions
\begin{eqnarray}
F_N&=&B\left( R\left( \theta ,\phi \right) ,\theta ,\phi \right)+\vec{B}\left( R\left( \theta ,\phi \right) ,\theta ,\phi \right) \cdot \hat{n}\left( \theta_{\rm d} ,\phi
\right)\label{pio1}\\
F_S&=&B\left( R\left( \theta ,\phi \right) ,\theta ,\phi \right)-\vec{B}\left( R\left( \theta ,\phi \right) ,\theta ,\phi \right) \cdot \hat{n}\left( \theta_{\rm d} ,\phi\right)\label{pio2}
\end{eqnarray}
with respect to $\theta_{\rm d}$ and $\phi$ ($\theta$ being understood as a function of $\theta_{\rm d}$, via Eq. (\ref{reu})). Here
\begin{equation}
R\left( \theta ,\phi \right) =\frac{AB}{\sqrt{A^2\sin^2\theta +B^2\cos^2\theta}}
\end{equation}
is the distance of the surface of the Earth to its center, and
\begin{equation}
\hat{n}\left( \theta_{\rm d} ,\phi \right) =\cos \theta_{\rm d} \cos \phi\, \hat{e}_{x}+\cos \theta_{\rm d} \sin \phi\, \hat{e}%
_{y}+\sin \theta_{\rm d}\, \hat{e}_{z}
\end{equation}
is the unit vector normal to the Earth's surface.

\section{Our team's project}\label{s5}

Our team's work was done in the context of the \textquotedblleft European Astro Pi Challenge Mission Space Lab 2020-21\textquotedblright , a contest promoted by ESA and the Raspberry Pi Foundation, where secondary school students are invited to write a code to be run on a Raspberry Pi device \cite{rasp} on board the International Space Station (ISS). Another example of work conducted by students in the Astro Pi Challenge, using the same equipment, can be found in \cite{mag}.

\begin{figure}[tbp]
\centering
\includegraphics[width=.75\textwidth]{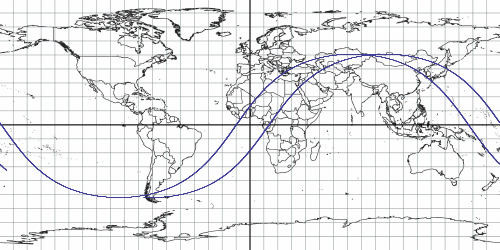}
\caption{The two orbits of the ISS mapped by our team.}
\label{fo}
\end{figure}

Our team of students (the last three authors of this article) wrote a code in Python to register the magnetic field at regular time intervals along the two orbits of the ISS during which the code was run (Fig. \ref{fo}). They recorded 4\,871 readings on 21 April 2021, from 02:24:23 GMT to 05:18:56 GMT, with an average interval between readings of 2.15 s.

\begin{figure}[tbp]
\centering
\includegraphics[width=.7\textwidth]{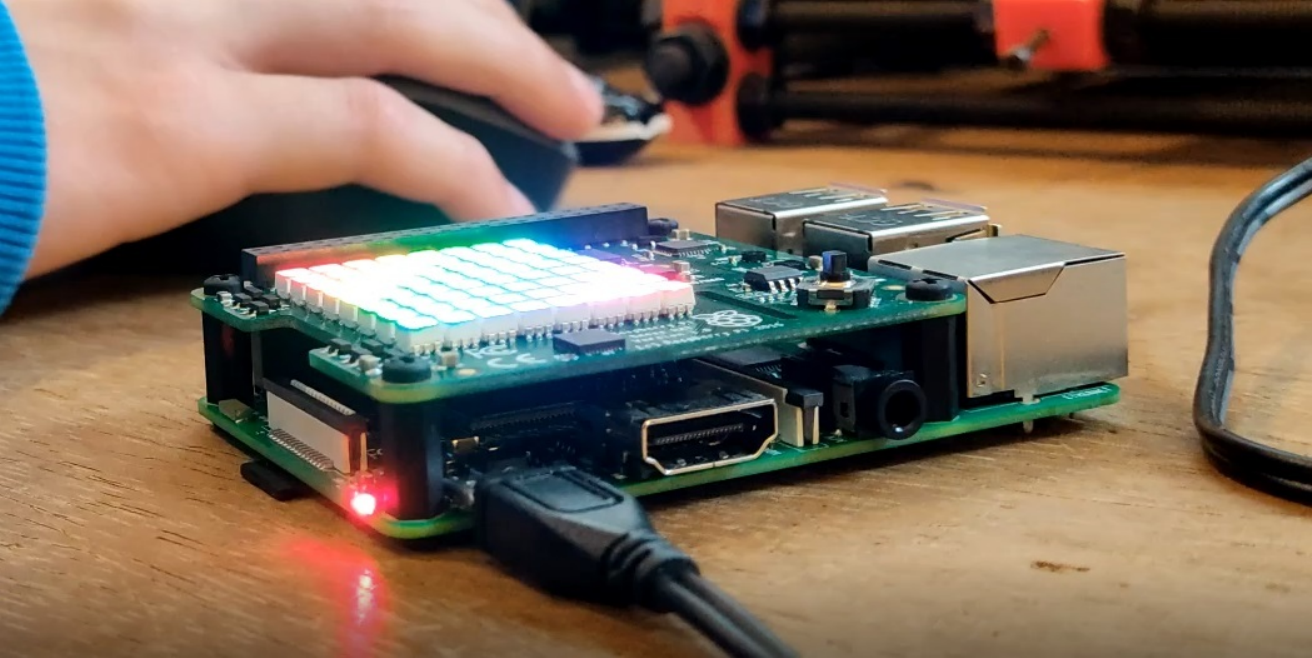}
\caption{The Raspberry Pi 3B computer together with the Sense Hat add-on board.}
\label{fig0}
\end{figure}

The magnetometer we used was the one included in the Sense Hat \cite{rasp}, an add-on board for the Raspberry Pi computer, containing a gyroscope, an accelerometer, a magnetometer, and sensors for temperature, pressure and humidity. The Raspberry Pi is an affordable single-board computer, costing approximately 80 \nolinebreak \euro, while the Sense Hat costs about 40 \nolinebreak \euro. The dimensions of the whole set (Raspberry Pi plus Sense Hat, Fig. \ref{fig0}) are approximately $9\times 6\times 2$ cm, and it weights less than 100 g. The Sense Hat magnetometer has a relatively low accuracy, of the order of hundreds of nT. The Sense Hat was made especially for the Astro Pi mission, but other low cost magnetometers can be attached to the computer, as was done in \cite{beg}.

The ISS circles the Earth with a 93 minute period at an average altitude of 400 km, and its orbit is inclined by $51.6^\circ$ with respect to the Equator. We had access to the ISS position coordinates for each measurement of the magnetic field, in the form of the altitude $h$, longitude $\phi$ and geodetic latitude $\theta_d$, and converted them to the cartesian
coordinates of the ECEF using Eqs. (\ref{piu1})-(\ref{piu2}).

However, we cannot use Eqs. (\ref{nu1})-(\ref{nu3}) to convert the components of the field between the ECEF frame and the magnetometer's frame because the magnetometer axes are not, in this case, aligned in the bottom-up, South-North and West-East directions. In fact, we were not provided with the information about the orientation of magnetometer inside the ISS. All we knew was that is was at rest with respect to the ISS. Therefore we first related the components of the magnetic field in the ECEF frame, $B_{x}$, $B_{y}$ and $B_{z}$, with its components in a moving frame fixed with respect to the ISS, $B_{\tilde{X}}$, $B_{\tilde{Y}}$ and $B_{\tilde{Z}}$, and then related, through a fixed but unknown rotation, the latter with the field components along the magnetometer axes, $B_{X}$, $B_{Y}$ and $B_{Z}$. The resulting transformations from the first to the third set of components replace Eqs. (\ref{nu1})-(\ref{nu3}) in our case. We note that the remaining results of Section \ref{s4} are unaffected and remain valid in our case.

\begin{figure}[tbp]
\centering
\includegraphics[width=.5\textwidth]{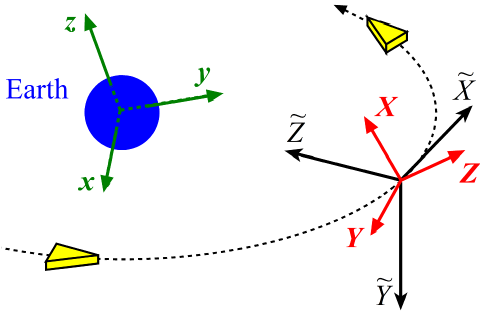}
\caption{The ECEF frame (lowercase letters), the ISS Local Vertical Local Horizontal (LVLH) reference frame (capital letters with a tilde) and the magnetometer's frame (capital letters), whose orientation we do not know.}
\label{fig2}
\end{figure}

The ISS is made to orbit the Earth with the same side permanently facing the Earth, and with the same side facing the direction of motion too (Fig. \ref{fig2}). This allows for an easy construction of a reference frame which is fixed with respect to the ISS. The LVLH (local vertical, local horizontal) reference frame of the ISS is defined in the following manner: the $\tilde{Z}$ axis points towards the Earth's center and the $\tilde{Y}$ axis is normal to the orbit, pointing in the opposite direction to the angular velocity; the frame is completed with the $\tilde{X}$ axis being orthogonal to the other two. Since the orbit of the ISS is approximately circular, the $\tilde{X}$ axis points in the direction of movement of the ISS.

In a fairly good approximation, we can write
\begin{eqnarray}
\hat{e}_{\tilde{X}} &=&\frac{\vec{r}_{i}\times \left( \vec{r}_{i}\times \vec{r}%
_{i-1}\right) }{r_{i}\left\Vert \vec{r}_{i}\times \vec{r}_{i-1}\right\Vert }=%
\frac{\left( \vec{r}_{i}\cdot \vec{r}_{i-1}\right) \vec{r}_{i}-r_{i}^{2}\vec{%
r}_{i-1}}{r_{i}\left\Vert \vec{r}_{i}\times \vec{r}_{i-1}\right\Vert } \\
\hat{e}_{\tilde{Y}} &=&\frac{\vec{r}_{i}\times \vec{r}_{i-1}}{\left\Vert \vec{r}%
_{i}\times \vec{r}_{i-1}\right\Vert } \\
\hat{e}_{\tilde{Z}} &=&-\frac{\vec{r}_{i}}{r_{i}}\ ,
\end{eqnarray}%
where $\vec{r}_{i}$ is the position of the ISS at a certain instant $t_{i}$ and $\vec{r}_{i-1}$ is its position at the instant $t_{i-1}$, immediately before $t_{i}$, all expressed in the ECEF reference frame.

Hence the components of the magnetic field in the LVLH reference frame are given by%
\begin{eqnarray}
B_{\tilde{X}} &=&\frac{\left( \vec{r}_{i}\cdot \vec{r}_{i-1}\right) \left(
\vec{B}_{i}\cdot \vec{r}_{i}\right) -r_{i}^{2}\left( \vec{B}_{i}\cdot \vec{r}%
_{i-1}\right) }{r_{i}\left\Vert \vec{r}_{i}\times \vec{r}_{i-1}\right\Vert }
\\
B_{\tilde{Y}} &=&\frac{\vec{B}_{i}\cdot \left( \vec{r}_{i}\times \vec{r}%
_{i-1}\right) }{\left\Vert \vec{r}_{i}\times \vec{r}_{i-1}\right\Vert } \\
B_{\tilde{Z}} &=&-\frac{\vec{B}_{i}\cdot \vec{r}_{i}}{r_{i}}\ ,
\end{eqnarray}%
where $\vec{B}_{i}$ is the expected magnetic field at instant $t_{i}$, expressed in the ECEF frame.

Since the magnetometer was fixed with respect to the ISS, the relation between these components and their counterparts in the magnetometer's frame $(B_{X}, B_{Y}, B_{Z})$ is simply given by a rotation:
\begin{equation}
\left[
\begin{tabular}{l}
$B_{X}$ \\
$B_{Y}$ \\
$B_{Z}$%
\end{tabular}%
\right] =M\left[
\begin{tabular}{l}
$B_{\tilde{X}}$ \\
$B_{\tilde{Y}}$ \\
$B_{\tilde{Z}}$%
\end{tabular}%
\right]\ , \label{cim}
\end{equation}%
with the rotation matrix $M$ depending on three unknown Euler angles $\alpha$, $\beta$ and $\gamma$:
\begin{equation}
M=\left[
\begin{tabular}{ccc}
$\cos \alpha \cos \beta $ & $\left(
\begin{array}{c}
\cos \alpha \sin \beta \sin \gamma\, - \\
-\sin \alpha \cos \gamma%
\end{array}%
\right) $ & $\left(
\begin{array}{c}
\cos \alpha \sin \beta \cos \gamma\, + \\
+\sin \alpha \sin \gamma%
\end{array}%
\right) $ \\
$\sin \alpha \cos \beta $ & $\left(
\begin{array}{c}
\sin \alpha \sin \beta \sin \gamma\, + \\
+\cos \alpha \cos \gamma%
\end{array}%
\right) $ & $\left(
\begin{array}{c}
\sin \alpha \sin \beta \cos \gamma\, - \\
-\cos \alpha \sin \gamma%
\end{array}%
\right) $ \\
$-\sin \beta $ & $\cos \beta \sin \gamma $ & $\cos \beta \cos \gamma $%
\end{tabular}%
\right]\ . \label{com}
\end{equation}%

In Eq. (\ref{cim}), the $B_{\left( X,Y,Z\right) }\left( g_{n}^{m},h_{n}^{m},\alpha ,\beta,\gamma \right) $ are the components of the magnetic field in the magnetometer frame and they are the ones to be used in Eq. (\ref{blaba}). Notice that these expressions depend on three extra parameters: the angles of the rotation matrix. This means that Eq. (\ref{blaba}) must be minimized for $\alpha $, $\beta $ and $\gamma $ too, though their values are of no interest for our purpose. This procedure added extra sources of error and more processing time to run the algorithms, which should be avoided, but could not within the rules of the \textquotedblleft Astro Pi Challenge\textquotedblright .

\section{Results}

\begin{figure}[tbp]
\centering
\includegraphics[width=.6\textwidth]{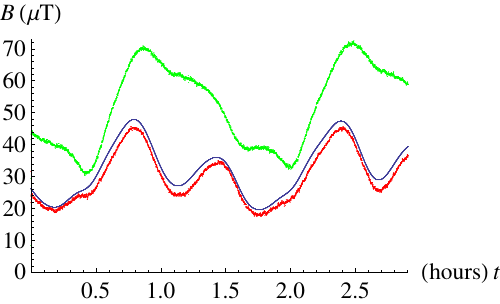}
\caption{Magnitude of the magnetic field as a function of time along the trajectory of the ISS. Upper curve: measured. Middle curve: expected from the ($N=13$) IGRF multipolar model. Lower curve: measured, corrected by subtracting a constant field.}
\label{fig}
\end{figure}

Although we were not provided with the orientation of the Raspberry Pi inside the ISS, we could at once compare the measured magnitude of the magnetic field $B=\sqrt{B_X^2+B_Y^2+B_Z^2}$ along the orbits of the ISS with the prediction from the IGRF model, since $B$ does not depend on the magnetometer's orientation. This is depicted in Fig. \ref{fig}.

Albeit agreeing overall in order of magnitude and in the position of the peaks, our data differed significatively from the estimate from the IGRF model. This could not be explained by variations of the Earth's magnetic field: changes on time scales greater than one year are of origin internal to the Earth, and are accounted for by the IGRF model, whereas seasonal variations are expected to be of the order of a few tens of nT \cite{ul,ula}. The lithospheric field, although of higher magnitude, should not exceed 20 nT at the ISS altitude \cite{sab}. Additional short term magnetic field variations may be due to ionospheric currents, or originate in the magnetosphere. In both cases, they should also occur on the scale of a few tens of nT \cite{ula,sab}, while we observe deviations reaching 10 $\mu $T, almost a thousand times greater!

This discrepancy could however come from the presence of a static magnetic field inside the ISS, or from an improper calibration offset, as both effects would shift the measurements by a constant field. In order to take this into account, we assumed the existence of such a field $\vec{b}^0$, with components $\left( b^0_X,b^0_Y,b^0_Z\right)$ in the magnetometer frame, subtracted it from the magnetometer readings $b_{X}^{i}$, $b_{Y}^{i}$ and $b_{Z}^{i}$, and, instead of minimizing Eq. (\ref{blaba}) we minimized
\begin{eqnarray}
&& S\left[ g_{l}^{m},h_{l}^{m},\alpha ,\beta,\gamma,b^0_X,b^0_Y,b^0_Z\right] =\sum_{i=1}^{P}\left[ \left( B_{X}^{i}%
\left[ g,h,\alpha ,\beta,\gamma\right] -b_{X}^{i}+b^0_X\right) ^{2}\right. \nonumber\\
&& \left. +\left( B_{Y}^{i}\left[ g,h,\alpha ,\beta,\gamma\right]
-b_{Y}^{i}+b^0_Y\right) ^{2}+\left( B_{Z}^{i}\left[ g,h,\alpha ,\beta,\gamma\right] -b_{Z}^{i}+b^0_Z\right)
^{2}\right]\label{blaba2}
\end{eqnarray}%
with respect to $g_{n}^{m}$, $h_{n}^{m}$, $\alpha$, $\beta$, $\gamma$, $b^0_X$, $b^0_Y$ and $b^0_Z$. Of course, just as when we introduced the Euler angles, having to find these extra parameters added further sources of error and processing time.

All calculations described in this article were performed using the \textit{Mathematica} computing software \cite{math}, and are available at the repository \cite{magaz}. Due to the presence of trigonometric functions in Eq. (\ref{com}), the minimization of $S$ produces several local minima. The built-in function \textit{NMinimize} attempts to find the global minimum but with no guarantee of success, as it turned out to happen in our case. We therefore minimized $S$ using the function \textit{FindMinimum}, which finds local minima starting from given seed values. The price to pay is that the program had to run for different seed values of the angles $\alpha $, $\beta $, and $\gamma $, until all the local minima were found. From these, the absolute minimum was determined.

As a measure of the accuracy with which our model fitted the measurements, we used the values of $g_{l}^{m}$, $h_{l}^{m}$, $\alpha$, $\beta$, $\gamma$, $b^0_X$, $b^0_Y$ and $b^0_Z$ obtained from the minimization process to compute the mean absolute error
\begin{equation}
\Delta B=\frac{1}{P}\sum_{i=1}^P\|\vec{B}_i\left( g,h,\alpha ,\beta,\gamma\right)-\vec{b}_i+\vec{b}^0_i\|\ .
\end{equation}
We obtained $\Delta B=1.9\ \mu$T for our data (and $\Delta B=3.6\ \mu$T for the data in Section \ref{sete}).

The intensity of the subtracted constant field $\vec{b}^0$ determined by the minimization process ($38.8\ \mu$T for our data and $22.0\ \mu$T for the data used in Section \ref{sete}) seemed too large either to be the average field inside the ISS or to result from a wrong calibration offset. It could also result from the magnetometer being close to magnetized or electronic equipment. We have no way to determine that, but we are satisfied that it must result from one of these possibilities, or from a combination of them, since its subtraction produced a very good match with the values predicted by the IGRF (as shown in Fig. \ref{fig}, where the lower curve is a plot of the absolute value of the vector $\left( b^i_X-b^0_X,b^i_Y-b^0_Y,b^i_Z-b^0_Z\right)$), and since other sources of distortion of the magnetic field would not have produced a constant shift.

\begin{table}[b!]
\centering
\begin{tabular}{|r|r|r|r|r|r|r|r|}
\hline
Moment & $g_1^0$ & $g_1^1$ & $h_1^1$ & $g_2^0$ & $g_2^1$ & $h_2^1$ & $g_2^2$ \\
\hline
IGRF & -29.40 & -1.44 & 4.62 & -2.51 & 2.97 & -3.03 & 1.67 \\
\hline
This work & -23.48 & -0.47 & 2.39 & -2.43 & 2.58 & -2.98 & 2.12 \\
\hline\hline
$h_2^2$ & $g_3^0$ & $g_3^1$ & $h_3^1$ & $g_3^2$ & $h_3^2$ & $g_3^3$ & $h_3^3$ \\
\hline
-0.76 & 1.37 & -2.39 & -0.07 & 1.24 & 0.24 & 0.51 & -0.54 \\
\hline
-0.85 & 2.12 & -2.51 & -2.74 & -0.08 & 0.22 & 0.27 & -0.28 \\
\hline
\end{tabular}
\caption{Multipolar moments obtained from our analysis and the IGRF reference values (in $\protect\mu$T).}
\label{taba1}
\end{table}

The values of the multipolar moments we obtained are given in Table \ref{taba1}, along with the IGRF's values for comparison. The latter were computed as
\begin{equation}
v\left( t\right) =v\left( 2020.0\right) +\dot{v}t\ ,
\end{equation}
where $v\left( 2020.0\right)$ is the value of any of the $g_{l}^{m}$ and $h_{l}^{m}$ for 1 January 2020, 00:00:00 GMT provided in \cite{al}, $\dot{v}$ is its expected linear variation for 2020-2025 \cite{al} and $t$ is the time elapsed between 2020.0 and the instant of the first reading of the magnetometer. Since we did not have any information on the precise location of our setup in the ISS, we could not assess the values for $\alpha$, $\beta$, and $\gamma$ that came out of the minimization.

\begin{table}[b!]
\centering
\begin{tabular}{||l||c|c|c||}
\hline
& IGRF & This work & Deviation \\ \hline\hline
NP & $86.4^\circ$N , $156.8^\circ$E & $70.4^\circ$N , $80.0^\circ$E & $19.1^\circ$ \\ \hline
SP & $64.0^\circ$S , $135.7^\circ$E & $57.5^\circ$S , $163.9^\circ$E & $15.1^\circ$ \\ \hline
\end{tabular}%
\caption{Location of Earth's dip poles obtained from our analysis and the IGRF values.}
\label{tabu1}
\end{table}

\begin{figure}[tbp]
\centering
\includegraphics[width=\textwidth]{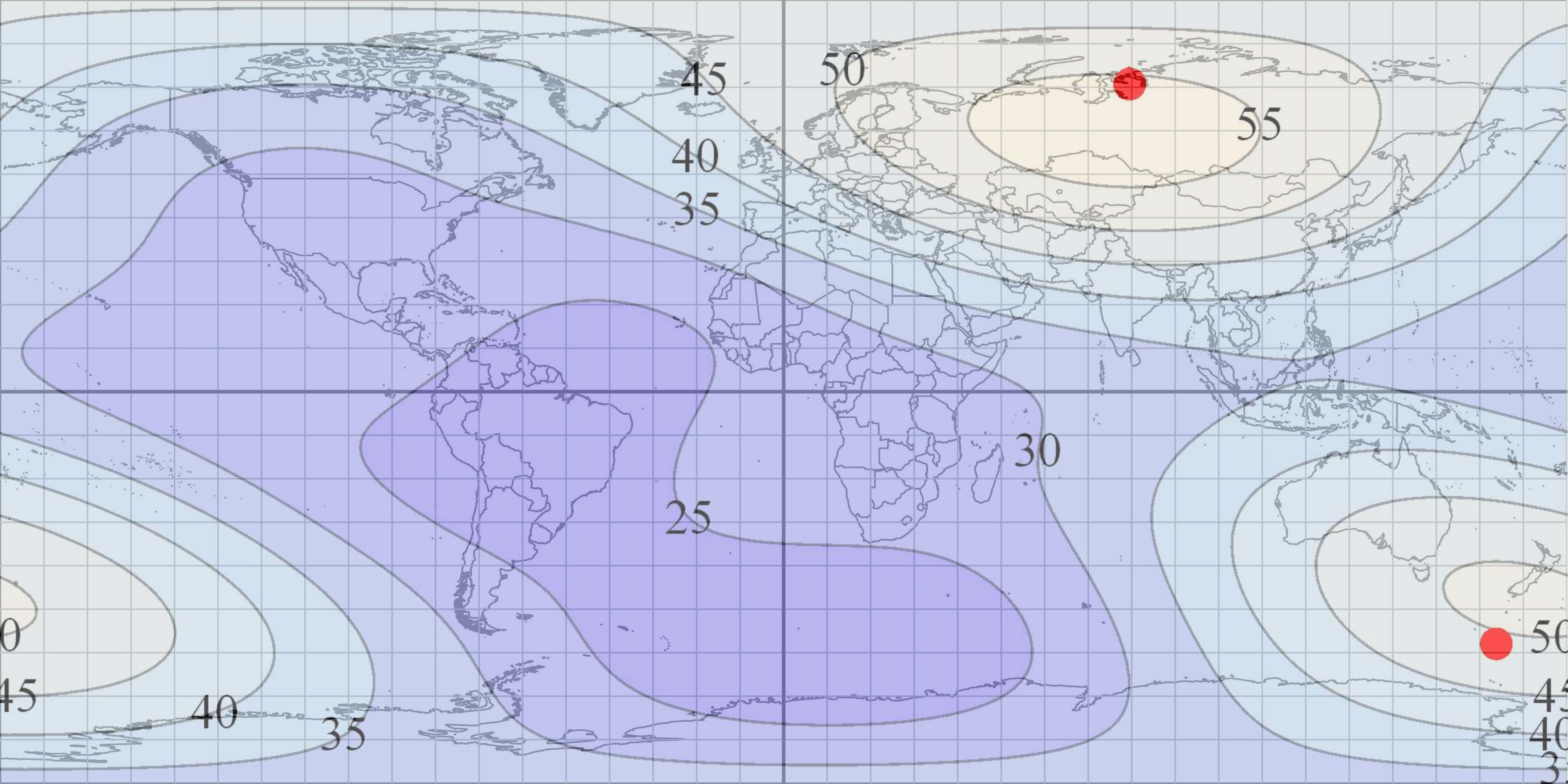}
\caption{Contour map for the magnitude of the magnetic field, using our results for the multipolar model to order 3. Separation between contour lines: 5 $\protect\mu $T. Darker areas represent weaker fields. The two small dots are the dip poles, as inferred from our measurements.}
\label{cp}
\end{figure}

The location of the magnetic dip poles was obtained by the method described in Section \ref{s4}, and is compared with the IGRF's values in Table \ref{tabu1}. Figure \ref{cp} shows a contour map of the Earth's magnetic field intensity, calculated from the values of the multipolar moments that we obtained.

All calculated multipolar moments agree with the ones from the IGRF both in order of magnitude and in sign. But they are not accurate. For example, the dipole moment ($l=1$) is $20.8\%$ weaker in magnitude than the expected value and $3.5^\circ$ off the right direction. The same lack of accuracy is visible in the calculated location of the magnetic poles (Table \ref{tabu1}) and in the intensity map in Fig. \ref{cp} (to be compared with the corresponding one in \cite{al}). Despite these uncertainties, some major features of the Earth's magnetic field, such as the South Atlantic anomaly, are visible.

\section{Another set of orbits}\label{sete}

\begin{figure}[tbp]
\centering
\includegraphics[width=.75\textwidth]{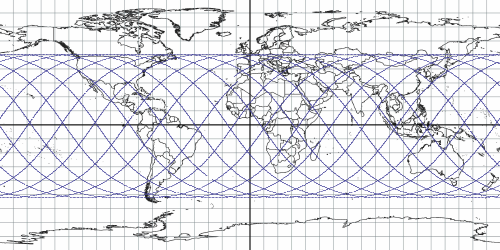}
\caption{The 15 orbits of the ISS extracted from the ``Raspberry Pi Learning Resources'' webpage.}
\label{fo2}
\end{figure}

Our results could hardly be improved by sampling more points along the probed orbits, as that would only augment the precision for shorter wavelengths, associated with higher order terms. However, they could benefit from a greater number of orbits. For that purpose, after the work for the \textquotedblleft Astro Pi Challenge\textquotedblright\ was completed, we repeated our analysis for a set of 15 orbits extracted from data provided by the Raspberry Pi site \cite{or}. This set of ISS orbits provides data that is more uniformly distributed along the surface of the Earth, forming a homogeneous grid between latitudes of $52^\circ$N and $52^\circ$S (Fig. \ref{fo2}), even though it has a lower number of measurements per orbit than our data. The selected data was recorded from 23 February 2016, 10:52:51 GMT, to 24 February 2016, 10:00:11 GMT. It comprised 8\,261 readings taken with an average interval between them of 10.08 s.

\begin{table}[b!]
\centering
\begin{tabular}{|r|r|r|r|r|r|r|r|}
\hline
Moment & $g_1^0$ & $g_1^1$ & $h_1^1$ & $g_2^0$ & $g_2^1$ & $h_2^1$ & $g_2^2$ \\
\hline
IGRF & -29.43 & -1.49 & 4.76 & -2.46 & 3.01 & -2.88 & 1.68 \\
\hline
15 orb. & -28.35 & -1.74 & 4.58 & -3.19 & 3.11 & -2.49 & 1.75 \\
\hline\hline
$h_2^2$ & $g_3^0$ & $g_3^1$ & $h_3^1$ & $g_3^2$ & $h_3^2$ & $g_3^3$ & $h_3^3$ \\
\hline
-0.66 & 1.35 & -2.36 & -0.11 & 1.23 & 0.24 & 0.57 & -0.54 \\
\hline
-0.49 & 1.55 & -2.35 & -0.49 & 1.07 & 0.52 & 0.67 & -0.41 \\
\hline
\end{tabular}
\caption{Multipolar moments obtained from our analysis of the 15 orbits and the IGRF reference values (in $\protect\mu$T).}
\label{taba2}
\end{table}

\begin{table}[b!]
\centering
\begin{tabular}{||l||c|c|c||}
\hline
& IGRF & 15 orb. & Deviation \\ \hline\hline
NP & $86.5^\circ$N , $167.8^\circ$W & $86.2^\circ$N , $33.7^\circ$E & $7.2^\circ$ \\ \hline
SP & $64.2^\circ$S , $136.4^\circ$E & $57.8^\circ$S , $138.8^\circ$E & $6.5^\circ$ \\ \hline
\end{tabular}%
\caption{Location of Earth's dip poles obtained from our analysis of the 15 orbits and the IGRF values.}
\label{tabu2}
\end{table}

The values of the multipolar moments we obtained are given in Table \ref{taba1}, along with the IGRF's values for comparison. The latter were computed for the instant of the first reading from a linear interpolation between the 2015.0 and the 2020.0 values in \cite{al}.

\begin{figure}[tbp]
\centering
\includegraphics[width=\textwidth]{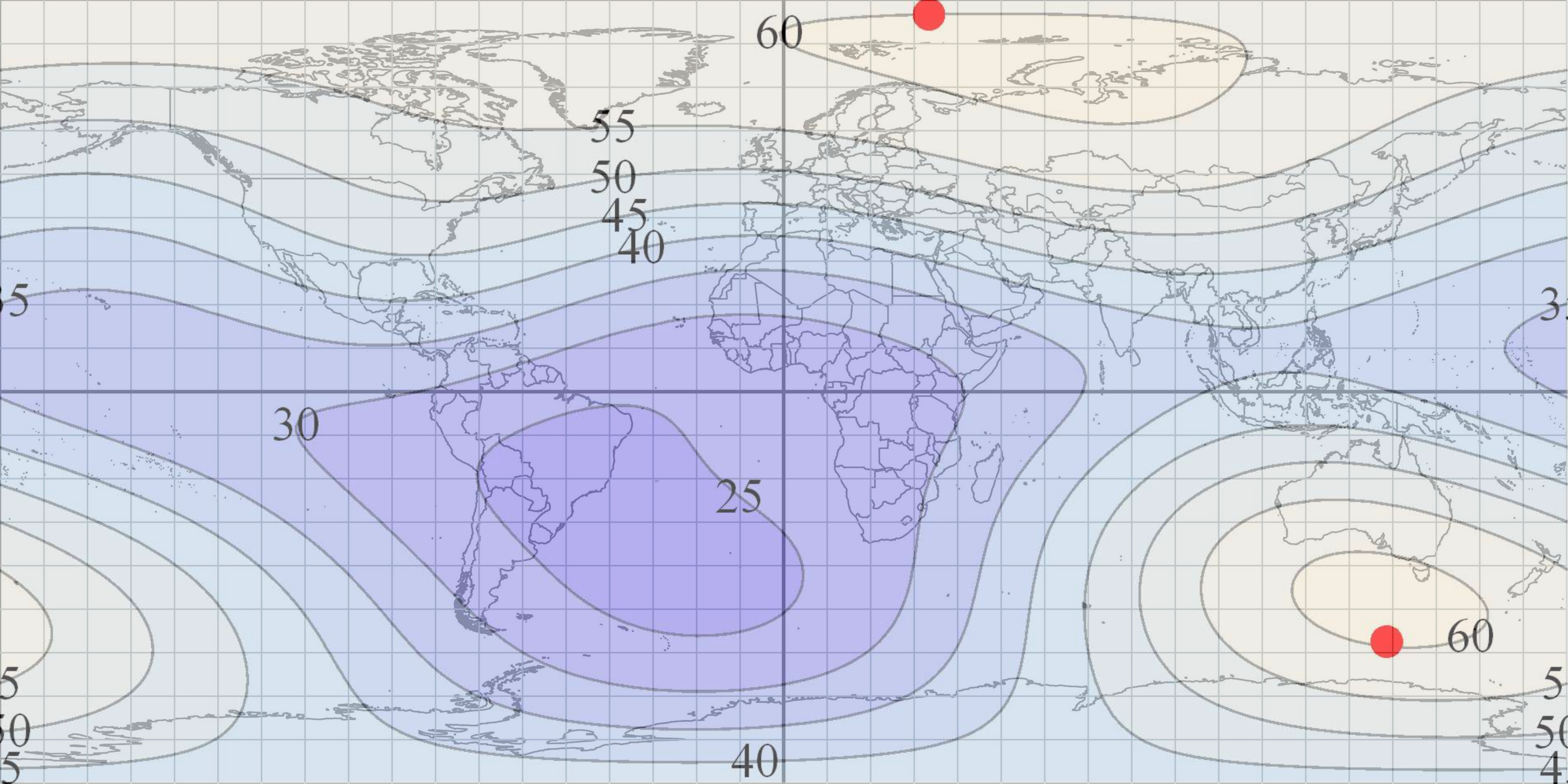}
\caption{Contour map for the magnitude of the magnetic field, using our analysis of the 15 orbits for the multipolar model to order 3. Separation between contour lines: 5 $\protect\mu $T. Darker areas represent weaker fields. The two small dots are the dip poles, as inferred in this analysis.}
\label{cp2}
\end{figure}

In Table \ref{tabu2} the location of the poles calculated from this set of orbits is compared with the IGRF's values. The location of the poles and the values of the momenta are not the same as the ones in Tables \ref{taba1} and \ref{tabu1}, due to the secular variation of the magnetic field between the two sets of measurements in 2016 and 2020. In Fig. \ref{cp2} we show the contour map of the expected field intensity for the values of the multipolar moments obtained for this set of 15 orbits.

We see that a better spatial sampling of the data provided great improvement in the determination of the multipolar moments (the dipole magnitude is only $3.6\%$ shorter of the IGRF value and its orientation is only $0.6^\circ$ off the IGRF axis) and the contour map in Fig. \ref{cp2} is much closer to the corresponding map in \cite{al}. However, the improvement in the determination of the location of the poles was more modest, probably due to the fact that the fields in the polar regions were still not sufficiently sampled.

\section{Conclusions}

Magnetic field is a concept in electromagnetism with which students have some familiarity because people are used to playing with magnets from an early age, and, by seeing the action of a compass, realize that the Earth itself has magnetic properties. The Earth's magnetic field thus provides an opportunity to create a bridge between the more abstract
and mathematical contents of electromagnetism courses with the students' day-to-day experience. The project reported here proposes such a bridge between Laplace's equation and its solutions in terms of spherical harmonics (a mathematical tool that also appears in more advanced disciplines of physics).

It came to us as a surprise that, with data points covering only a thin ring over the surface of the Earth, and 3 hours of measurement with a low-cost magnetometer onboard the ISS, one could reconstruct the main features of the Earth's magnetic field: its magnitude, its angle with respect to the planet's rotation axis, its non-uniformity, and even some of its anomalies. We made use of a Raspberry Pi-powered magnetometer on board the ISS, but the project could easily be adapted to ground measurements with state-of-the-art gaussmeters, Arduino-powered Hall sensors, or even smartphone magnetometers, and take advantage of the internet or social media to devise a participatory science experiment.

\section*{Acknowledgments}

We thank Ronald Merrill, Erwan Th\'{e}bault and Yosuke Yamazaki for
providing clarifications, and Maria Beatriz Cachim for support provided at
Escola Secund\'{a}ria Domingos Rebelo. We would also like to thank the reviewers of the American Journal of Physics for their helpful remarks which greatly contributed to the improvement of this article.


\begin{thebibliography}{99}
\bibitem{jac} J. Jackson, \textit{Classical Electrodynamics}, Wiley (1999).

\bibitem{bar} D. Bartlett, \textit{Essentials of Positioning and Location Technology}, Cambridge UP (2013).

\bibitem{sp} D. Stump and G. Pollack, \textquotedblleft A current sheet model for the Earth's magnetic field\textquotedblright , Am. J. Phys. \textbf{66}, 802-810
(1998).

\bibitem{md} H. Meyers and W. Davis, \textquotedblleft A Profile of the Geomagnetic Model User and Abuser\textquotedblright , J. Geomag. Geoelectr. \textbf{42}, 1079-1085 (1990).

\bibitem{mer} R. Merrill, M. McElhinny and P. McFadde, \textit{The magnetic
Field of the Earth}, Academic Press (1996).

\bibitem{al} P. Alken et al, \textquotedblleft International Geomagnetic Reference
Field: the thirteenth generation\textquotedblright , Earth Planets and Space, \textbf{73}, 49
(2021).

\bibitem{swarm} The ESA Swarm mission is currently the primary satellite data source (\verb|https://earth.esa.int/eogateway/missions/swarm|).

\bibitem{wmm} A. Chulliat, W. Brown et al, \textquotedblleft The US/UK World Magnetic Model for 2020-2025: Technical Report\textquotedblright , National Centers for Environmental Information, NOAA (2020).

\bibitem{inter} \verb|http://www.intermagnet.org/|

\bibitem{beg} C. Beggan and S. Marple, \textquotedblleft Building a Raspberry Pi school magnetometer network in the UK\textquotedblright , Geosci. Commun. \textbf{1}, 25-34 (2018).

\bibitem{hof} B. Hofmann-Wellenhof, H. Lichtenegger and J. Collins, \textit{GPS - Theory and Practice}, Springer (1994).

\bibitem{rasp} \verb|https://www.raspberrypi.org/|

\bibitem{mag} T. Magalh\~{a}es et al., \textquotedblleft Observation of atmospheric
gravity waves using a Raspberry Pi camera module on board the International
Space Station\textquotedblright , Acta Astronautica \textbf{182}, 416-423 (2021).

\bibitem{ul} S. Malin and D. Winch, \textquotedblleft Annual variation of the
geomagnetic field\textquotedblright , Geophys. J. Int. \textbf{124}, 170-174 (1996).

\bibitem{ula} V. Courtillot and J. Le Mou\"{e}l, \textquotedblleft Time variations of
the Earth's magnetic field: From daily to secular\textquotedblright , Ann. Rev. Earth Planet.
Sci. \textbf{16}, 389-476 (1988).

\bibitem{sab} T. Sabaka, N. Olsen and R. Langel, \textquotedblleft A comprehensive model of the quiet-time, near-Earth
magnetic field: phase 3\textquotedblright , Geophys. J. Int. \textbf{151}, 32-68 (2002).

\bibitem{math} \verb|https://www.wolfram.com/mathematica/|

\bibitem{magaz}
\verb|https://github.com/magnetometer/iss/releases/tag/issmag|

\bibitem{or}
Astro Pi Flight Data Analysis, by the Raspberry Pi Foundation:
\verb|https://github.com/raspberrypilearning/astro-pi-flight-data-|
\verb|analysis/raw/master/data/Columbus_Ed_astro_pi_datalog.csv.zip|
\end{thebibliography}
\end{document}